# First experimental proof of PET imaging based on multi-anode MCP-PMTs with Cherenkov radiator-integrated window

**Weiyan Pan[1,2,3,4], Lingyue Chen[1,2], Guorui Huang[5], Jun Hu[1,2], Wei Hou[5], Xianchao Huang[1,3,4*], Xiaorou Han[1,3,4], Xiaoshan Jiang[1,2], Zhen Jin[5], Daowu Li[1,3,4], Jingwen Li[5], Shulin Liu[1], Zehong Liang[1], Lishuang Ma[1], Zhe Ning[1,2], Sen Qian[1], Ling Ren[5], Jianning Sun[5], Shuguang Si[5], Yunhua Sun[1,2], Long Wei[1,2,3,4], Ning Wang[5], Qing Wei[1,2,3,4], Qi Wu[1,2], Tianyi Wang[1,2,3,4], Xin Wang[1,2], Xingchao Wang[5], Yangfu Wang[1,2,6], Yifang Wang[1,2], Yingjie Wang[1,2,4,5], Zhi Wang[5], Hang Yuan[1,2,4,5], Jingbo Ye[1,2], Xiongbo Yan[1,2], Meiling Zhu[1,3,4], Zhiming Zhang[1,2,3,4], On behalf of the MCP-PMT workgroup**

[1] Institute of High Energy Physics, Chinese Academy of Sciences, Beijing 100049, China
[2] University of Chinese Academy of Sciences, Beijing 100049, China
[3] Jinan Laboratory of Applied Nuclear Science, Jinan 250131, China
[4] CAEA Centre of Excellence on Nuclear Technology Applications for Nuclear Detection and Imaging. Beijing, China
[5] North Night Vision Science & Technology (Nanjing) Research Institute Co. Ltd, Nanjing 211100, China
* Corresponding author, E-mail: huangxc@ihep.ac.cn

## Abstract

*Objective.* Achieving a higher signal-to-noise ratio (SNR) gain or direct positron emission imaging (dPEI) by improving the coincidence time resolution (CTR) of time-of-flight positron emission tomography (TOF-PET) systems is paramount for many advanced clinical PET imaging applications. This places greater demands on the timing performance of all PET system components.
*Approach.* An effective method for enhancing detector time resolution is to use microchannel plate photomultiplier tubes (MCP-PMTs) for prompt Cherenkov photon detection. In this study, we developed a dual-module PET imaging experimental platform.
This platform uses two 8 × 8-anode Cherenkov radiator-integrated window MCP-PMTs, which was developed in-house. It also uses multi-channel electronics system, which was also developed in-house. We designed specific calibration and correction methods for the platform to meet the experimental requirements.
*Main results.* Based on this dual-module platform, a total FWHM of 203.6 ps (σ = 86.5 ps) was achieved. Multi-component analysis of the time spectrum was performed based on the interaction positions of annihilation photon pairs in different coincidence events. Imaging experiments were conducted using various types of radiation sources and phantoms. Spatial resolution was evaluated, confirming the platform's ability to distinguish 4 mm spots in Derenzo-like phantoms. Furthermore, preliminary improvements in image quality were verified when incorporating TOF information into the dual-module platform.
*Significance.* In this study, we achieved the first experimental verification of module-level PET imaging based on the detection of Cherenkov light. This has overcome the limitations of single-pixel detectors in previous related research and reduced the radiation dose and acquisition time required for imaging by nearly 10 times. Additionally, this study evaluated the potential of a dual-module platform for achieving higher temporal performance and direct imaging, elucidated the importance of developing $PbF_2$ integrated window multi-anode MCP-PMTs, and provided experimental evidence for future performance enhancements in Cherenkov light detection-based PET systems.

Keywords: TOF-PET; Multi-anode MCP-PMT; Cherenkov radiation;

## 1. Introduction

Positron emission tomography (PET) is a valuable molecular imaging technology characterized by its unique functional imaging properties. It plays a crucial role in drug research and disease diagnosis, particularly in the diagnosis of cancer (Boellaard *et al* 2015, Schwenck *et al* 2023) and various neurodegenerative diseases (Cavaliere *et al* 2020, Smith *et al* 2023). As clinical and research applications expand, new uses for PET have emerged. These include immuno-PET imaging (Lugat *et al* 2022), infectious disease imaging (Eibschutz *et al* 2022), and fetal imaging (Smith *et al* 2024), which require improved PET system sensitivity for faster and lower-dose imaging.

One of the principal factors influencing the signal-to-noise ratio (SNR) and image quality in a PET system is the time resolution (Karp *et al* 2008). Time-of-flight PET (TOF-PET) imaging incorporates TOF information into the reconstruction process. The utilization of TOF information in conjunction with the time resolution of the system restricts the reconstructed locations of the annihilation point to a specific range, as opposed to the equal weighting of all locations along the entire line of response (LOR) as in traditional PET. This results in a gain in SNR and enhances the effective sensitivity of the PET system (Lecoq 2017). As a new imaging modality that can directly localize the annihilation position using TOF information without tomographic reconstruction, direct positron emission imaging (dPEI) (Kwon *et al* 2021a) brings greater performance enhancements to the PET system, enables more imaging methods and imaging algorithms (Onishi *et al* 2024), and puts higher demands on the fast-timing detector (Lecoq *et al* 2020).

Currently, the combination of lutetium-based scintillators and silicon photomultipliers (SiPMs) represents the prevailing detector scheme for PET systems (Gundacker *et al* 2020, Nadig *et al* 2023, Pourashraf *et al* 2024). The coincidence time resolution (CTR) of the most advanced PET systems has reached approximately 200 ps Full Width at Half Maximum (FWHM) (Van Sluis *et al* 2019, Prenosil *et al* 2022), with the best systems achieving 178 ps FWHM (Siemens Healthineers n.d.). At this level of coincidence time resolution, the benefits of TOF technology are already evident. However, there is still a gap to be bridged before truly effective dPEI can be achieved or many of the advanced PET applications mentioned above can be realized. Therefore, further improvements are required in the timing performance of the PET detector.

To further improve the timing performance of PET detectors, it is imperative to minimize the time uncertainty derived from the photoelectric sensor throughout the photoelectric conversion and multiplication process. This necessitates a reduction in the single-photon time resolution (SPTR) (Gundacker *et al* 2023, 2019) or the Single photoelectron transit time spread ($TTS_{@SPE}$) (Bortfeldt *et al* 2020, Li *et al* 2020) of the photoelectric sensor. With regard to crystal luminescence, it is possible to develop scintillation crystals with a higher light yield and faster emission (Gundacker *et al* 2016, 2018). Furthermore, the constraints of the scintillation luminescence process itself can be addressed by exploring alternative luminescence mechanisms. Consequently, another effective approach for improving the time resolution of PET detectors is to utilize faster luminescence mechanisms for detecting annihilation gamma rays. It is anticipated that this improvement will be significant (Korpar *et al* 2011). Cherenkov radiation is a phenomenon that occurs when a charged particle, such as an electron, travels through a dielectric medium at a speed exceeding the phase velocity of light in that medium (Kobzev 2010). Cherenkov photons are generated through electromagnetic polarization and depolarization of the medium, exhibiting a ‘prompt’ characteristic. Therefore, the utilization of Cherenkov light can serve to mitigate the time uncertainty that is inherent to the distribution of photons. This time uncertainty is relatively pronounced during the scintillation process. Consequently, Cherenkov photons can provide more precise timestamps than scintillation photons (Han *et al* 2021), and PET detection using Cherenkov light is considered a promising way to achieve better CTR.

However, it is worth mentioning that the energy of the annihilation γ-rays is 511 keV, which results in a relatively low yield of Cherenkov photons (Kratochwil *et al* 2021). Furthermore, due to photon losses occurring during the transport and reception of light, Cherenkov photo detection in PET application often results in the detection of only a few photons or even a single photon. The microchannel plate photomultipliers (MCP-PMTs) employ a microchannel plate structure to replace the discrete multiplication electrode in traditional PMTs, thereby achieving excellent timing performance. Meanwhile, MCP-PMTs preserve the low-noise capability for single-photon detection inherent to conventional PMTs, in contrast to SiPMs. These characteristics are highly compatible with the properties of Cherenkov photons, which are few in number and have fast responses, making it an optimal choice for high time resolution in PET detection using Cherenkov light.

The timing performance of PET detection using Cherenkov light has been verified and reported. In one study, the Cherenkov radiator was used as the window material for the single-anode MCP-PMT and a CTR of 30.1 ps FWHM was obtained(Ota *et al* 2019a). This choice aims to reduce the transmission loss of Cherenkov photons at the interface. In a follow-up study, the CTR was further enhanced to 26.4 ps

FWHM , and dPEI without reconstruction was achieved for the first time (Kwon *et al* 2021b). However, because the MCP-PMT used in the study was single-anode, it lacked positional resolution. As a result, direct imaging must be achieved by collimated translational scanning. In addition, its sensitive area and overall shape prevent high-density multimodule assembly, making it difficult to meet the efficiency requirements of large PET systems. Therefore, the square multi-anode MCP-PMT is more suitable for PET applications due to its position-sensitive characteristics and ability to assemble multiple modules. Consequently, the square multi-anode MCP-PMT has been the focus of some research (Ma *et al* 2023, Jones *et al* 2023) and serves as the detector solution used in this study.

In this work, we completed the experimental setup of a dual-module PET imaging platform using a pair of 8 × 8-anode Cherenkov radiator-integrated window MCP-PMTs specifically designed and developed by our team. The readout and timing measurement of the MCP-PMT Cherenkov signals were accomplished using a multi-channel discriminator-based application specific integrated circuit (ASIC) chip, combined with a self-developed multi-channel time-to-digital converter (TDC). Calibration and related corrections for each component of the platform were performed based on the characteristics of the multi-anode MCP-PMT and multi-channel electronics. A CTR test of the entire system was performed, resulting in a measured CTR of 203.6 ps FWHM. Imaging experiments were also carried out using point and rod sources, as well as a custom-designed '$e^+$' phantom with specific structures, achieving module-level PET imaging based on Cherenkov light detection for the first time.

## 2. Materials

### *2.1 Multi-anode Cherenkov Radiator-integrated Window MCP-PMTs*

In this work, we employed self-developed multi-anode Cherenkov radiator-integrated window MCP-PMTs for annihilation gamma-ray detection based on the Cherenkov mechanism, as shown in Figure 1a. The Cherenkov radiator-integrated window MCP-PMT uses lead glass as the window material, aiming to minimize light loss at the interface. The main parameters of the lead glass material used for the Cherenkov radiator integrated window multi-anode MCP-PMT are shown in Table 1.

**Table 1. Summary of parameters of lead glass window material used in Cerenkov radiator integrated window MCP-PMT.**

| Parameter | Lead glass (RefractiveIndex.INFO n.d.) |
|---|---|
| Compound composition | 75% PbO+25%SiO2 |
| Density g/mm$^3$ | 5.49 |
| Refractive index n (365-1014 nm) | 1.81-1.95 |
| Effective atomic number $Z_{eff}$ | 62.5 |
| Cut-off wavelength nm | 370 |
| Threshold kinetic energy for Cherenkov radiation (electron) keV | 85.3-100.2 |

As shown in Table 2, the CTR and detection efficiency for different lead glass material thicknesses were simulated in this study using Geant4, based on the above parameters.

**Table 2. Simulation results of CTR and detection efficiency for different Cherenkov radiator thicknesses.**

| | Thickness | 2 mm | 3 mm | 4 mm | 5 mm | 7 mm | 10 mm |
|---|---|---|---|---|---|---|---|
| TTS$_{@SPE}$: 70.65 ps FWHM (30 ps σ) | CTR (FWHM)/ ps | 88.7 | 91.1 | 93.2 | 94.5 | 100.2 | 133.8 |
| | Detection efficiency | 4.9% | 7.0% | 8.9% | 10.5% | 14.0% | 17.6% |

Considering the detection efficiency, timing performance, and manufacturing process requirements, the window material thickness was determined to be 5.5 mm. As shown in Figure 1b, the anodes of the MCP-PMT in the Cherenkov radiation detector are arranged in an 8 × 8 array with a pixel size of 5.5 mm × 5.5 mm and a gap of 0.3 mm. The sensitive area of the entire multi-anode Cherenkov radiator-integrated window MCP-PMT is 46.1 mm × 46.1 mm. Typical performance parameters of the multi-anode Cherenkov radiator-integrated window MCP-PMT were obtained by testing the central anode of the multi-anode MCP-PMT, as shown in Table 3.

**Table 3. Typical performance parameters of multi-anode Cherenkov radiator-integrated window MCP-PMTs.**

| Detector | HV | Gain | Amplitude$_{@SPE}$ | Rise time | Fall time | TTS$_{@SPE}$ |
|---|---|---|---|---|---|---|
| MCP-PMT Sample 1 | -1440 V | 3.9×10$^6$ | 56.4 mV | 296 ps | 431 ps | 64.5 ps FWHM (27.4 ps σ) |
| MCP-PMT Sample 2 | -1485 V | 4.4×10$^6$ | 52.4 mV | 287 ps | 479 ps | 67.4 ps FWHM (28.6 ps σ) |

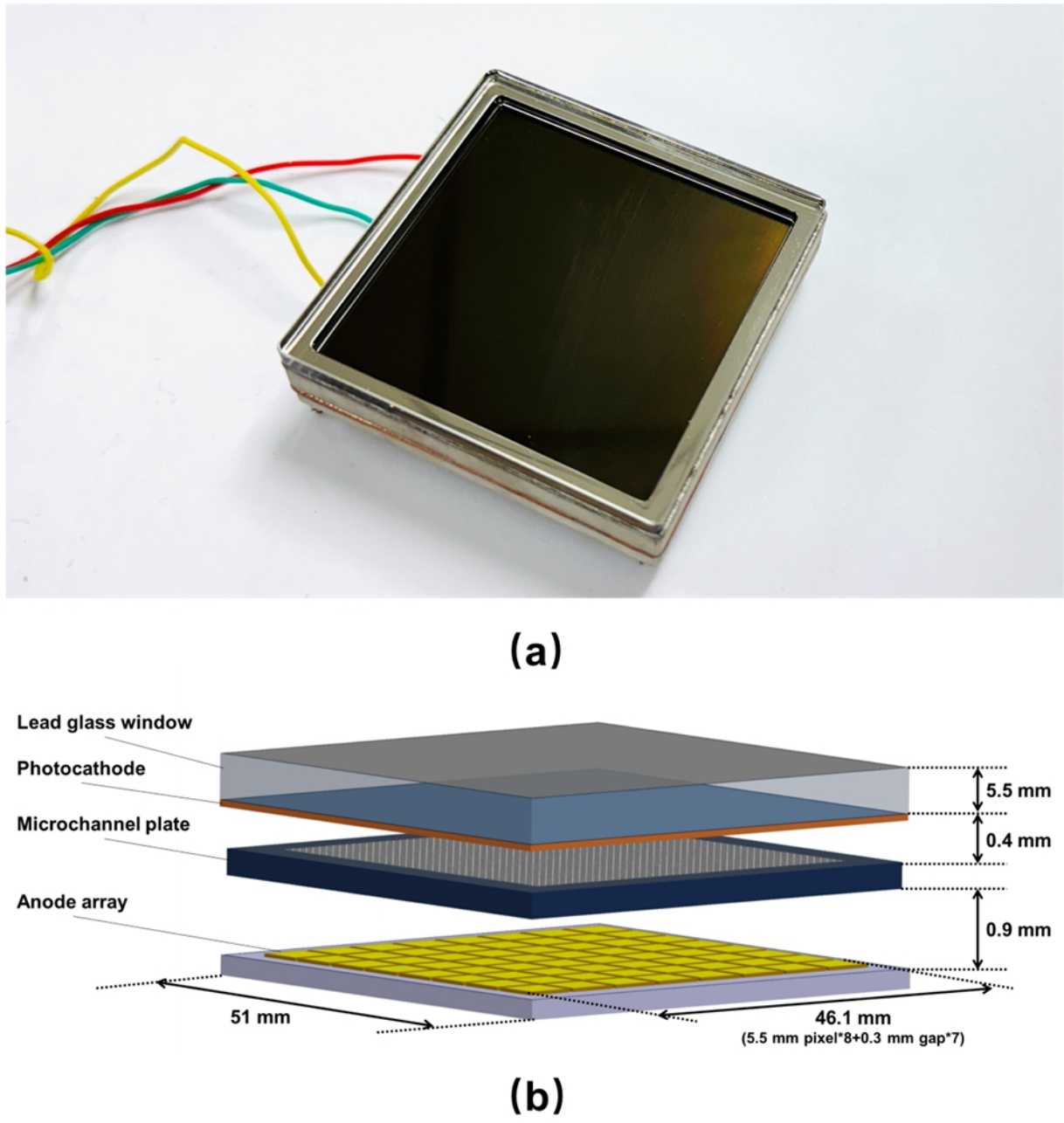

**Figure 1. Photo of the multi-anode Cherenkov radiator-integrated window MCP-PMTs(a) and a schematic diagram of its structure(b).**

## *2.2 Electronics*

The Cherenkov signal output from the Cherenkov radiator-integrated window MCP-PMT has a relatively small amplitude and a much faster rise time of less than 300 ps. To address these characteristics of the Cherenkov signal, our team designed a multi-channel high-speed discriminator ASIC called FPMTread, the internal schematic diagram is shown in Figure 2a. The FPMTread chip pre-amplifies the input signal through the receiver stage and adds it on a baseline voltage, which is formed through a resistor divider inside the chip and is about 700mV, which is there to help the chip achieve a larger input dynamic range. Subsequently, the signal after adding the baseline is compared to a threshold, and amplifies the part of the signal that exceeds the threshold to saturation. It is then differentially output through the Current mode logic (CML) driver, and the width of the output signal is related to the over-threshold time of the original signal. The schematic diagram of input and output signals is shown in Figure 2b. The performance parameters of the chip are shown in Table 4, and it is also compared with other discriminator ASICs currently used in PET.

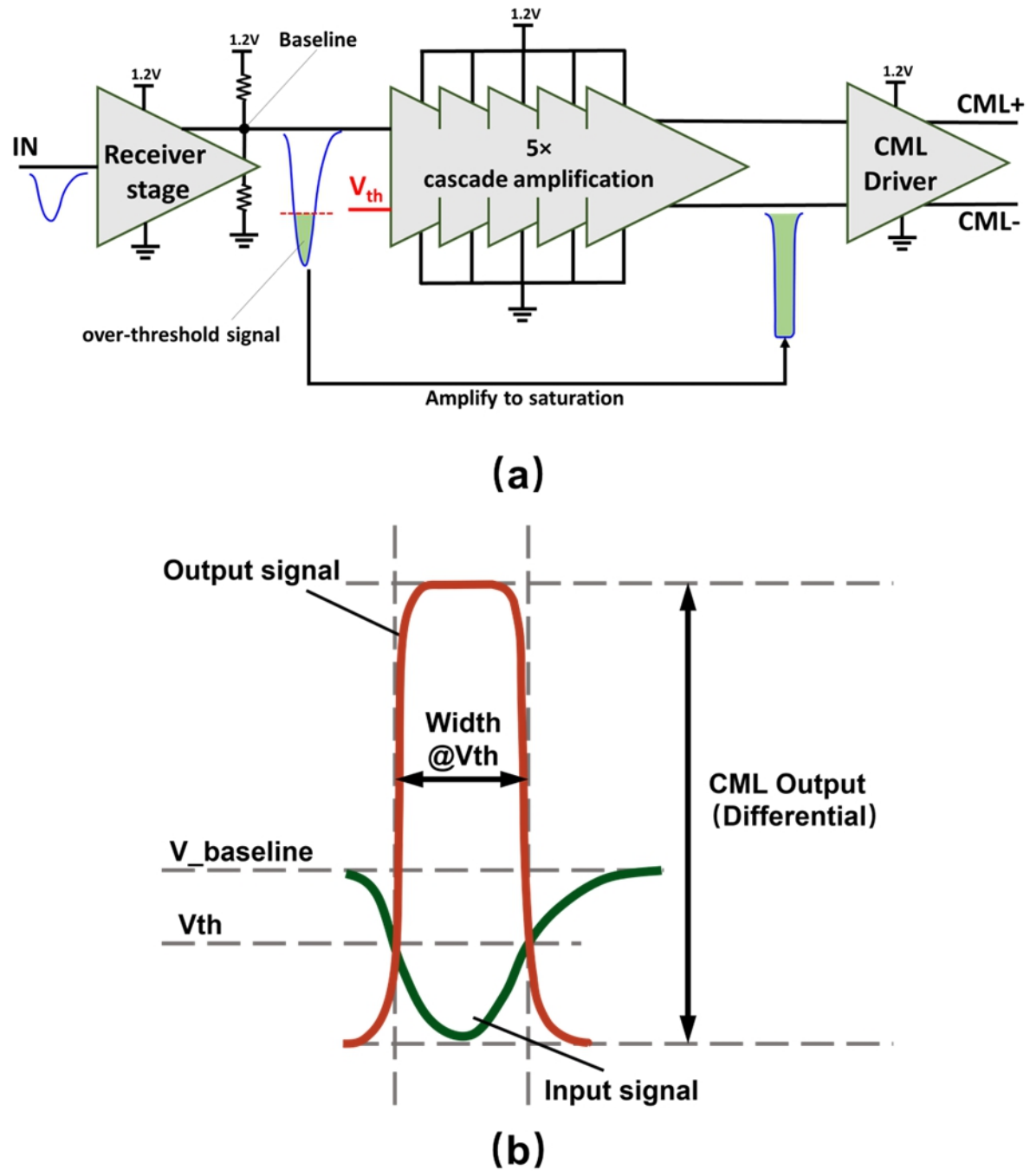

**Figure 2. The internal schematic diagram of FPMTread ASIC chip (a) and schematic diagram of input and output signals (b).**

**Table 4. The main performance parameters of the FPMTread and comparison with mainstream multi-channel discriminator ASICs.**

| Organization | IHEP | CERN | University of Barcelona |
|---|---|---|---|
| ASIC | FPMTread | NINO (Anghinolfi *et al* 2004) | HRflexTOT (Sanchez et al 2022) |
| Technology | 0.13 µm CMOS | 0.25 µm CMOS | 0.18 µm CMOS |
| Bandwidth | 2 GHz | 500 MHz | 500 MHz |
| Number of channels | 4 | 8 | 16 |
| Power consumption | 45 mW/ch | 30 mW/ch | 11 mW/ch |
| Time jitter | 10.9 ps FWHM (4.62 ps σ) | <58.9 ps FWHM (25 ps σ) | 16.5 ps FWHM (7 ps σ) |

An FPGA-based TDC is used as the digital electronics to implement the measurement of time information. The board has 64 TDC channels, with every two channels forming a group responsible for detecting the time information of the rising and falling edges of a signal to achieve the measurement of the pulse width. Each TDC board can realize the time measurement of 32 detector channels. The TDC has a measured precision of 4.77 ps RMS on average and is capable of achieving multi-module synchronous measurement(Liang *et al* 2024).

In order to meet the integration requirements for multi-channel measurements and to reduce the performance loss due to signal transfer and transmission, we combined the discriminator ASIC readout circuits and the FPGA-based TDC circuits into one electronic board. Each electronic board contains eight 4-channel FPMTread discriminator ASICs and one 32-channel FPGA-based TDC circuit. Independent threshold adjustment of each discriminator channel is also implemented by the FPGA in combination with multi-channel digital-to-analogue converter (DAC) chips. Differential clock ports for receiving synchronous clocks and an RST port for receiving reference signals have also been added for multi-module time synchronization. Based on the test method shown in Figure 3a, the electronic board measured a timing precision of 10.34 ps (σ) under the condition of multi-module synchronization. The measurement results are shown in Figure 3b.

### *2.3 Detector Modules*

To integrate the electronics and the Cherenkov radiator-integrated window MCP-PMT into a detector module, a readout board was designed to transmit the signals from the 64 anodes using coaxial connectors. To ensure the quality and integrity of the signal transmission, the signal was fed into the electronics board via a 10 cm coaxial cable. The crosstalk between the channels of the integrated electronics board was evaluated, and it was verified that the crosstalk level of adjacent channels was <1% after a 500 MHz frequency sine wave signal was fed into the electronics channels, which basically proved that the crosstalk would not affect the time measurement process. A metal structure was designed to support and connect the electronics boards to the Cherenkov radiator-integrated window MCP-PMT, allowing effective heat dissipation. The complete detector module is shown in Figure 3c. Each module consists of an 8 × 8 anode Cherenkov radiator-integrated window MCP-PMT and two 32-channel electronics boards.

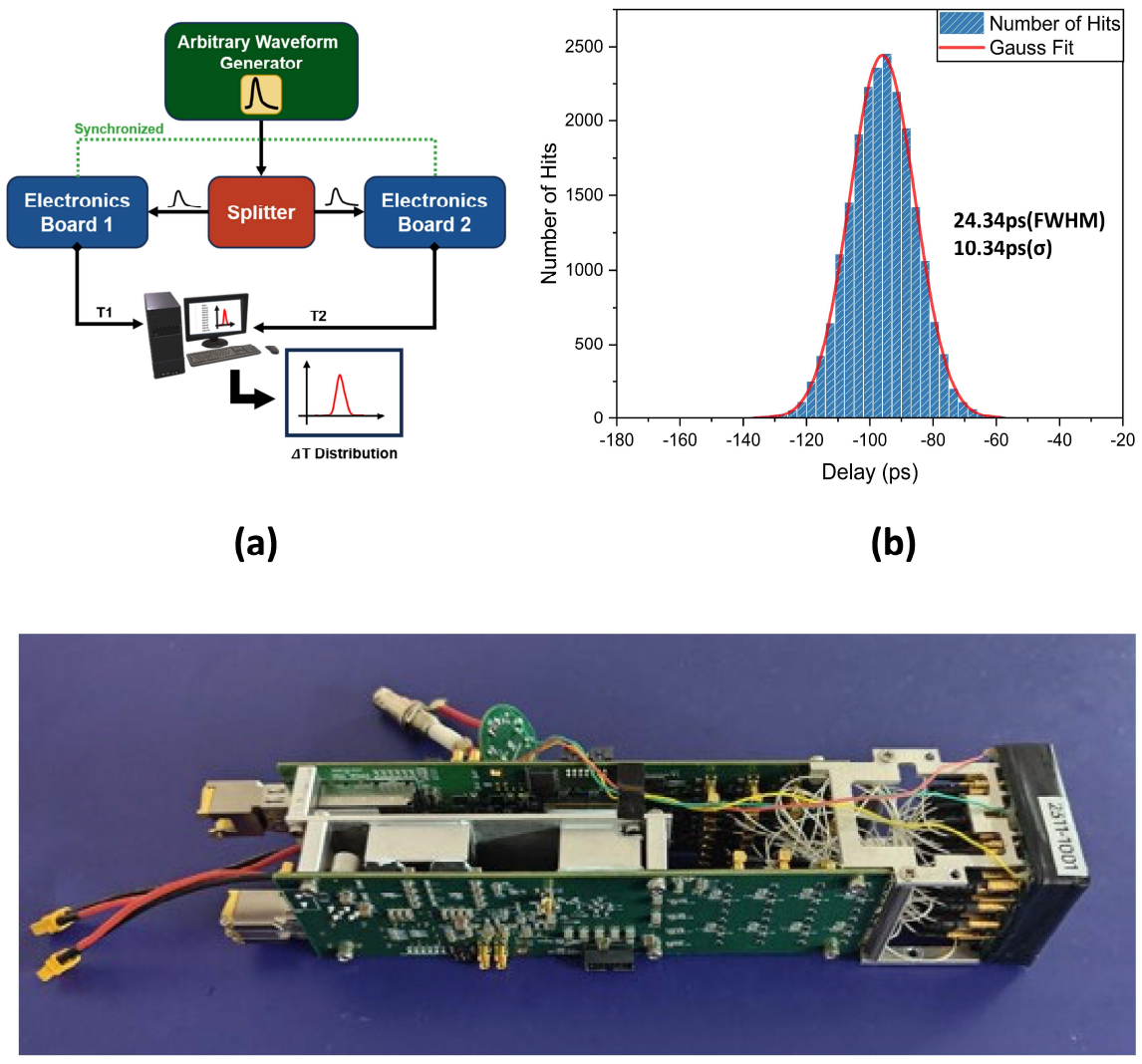

**Figure 3. Schematic diagram of the method for the timing performance test of electronics (a) and the test results (b). Detector module based on Cherenkov radiator-integrated window MCP-PMT (c).**

## 3. Methods

### *3.1 Experimental Setup of Dual-module PET Imaging Platform*

Figure 4 shows the experimental setup for a dual-module PET imaging platform based on Cherenkov radiator-integrated window MCP-PMTs. Two detector modules were arranged face to face at a distance of 46.1 mm, corresponding to the side length of the detector sensitivity region. The radioactive source or imaging object used in the experiment was placed in the center of the field of view (FOV) and was equipped with a rotary stage for multi-angle acquisition. After calibrating the voltage-count rate curve, the voltages of the two MCP-PMTs were set to -1500 V and -1550 V (V6534 HV power supply module, CAEN). A commercial clock board (Si5361-EB, Skyworks) and a specifically designed reset board form a clock and synchronization module. The commercial clock board supplies synchronized clocks to multiple modules. The reset board supplies reference signals with fixed time differences to all modules based on the synchronized clock. This signal clears the coarse time of the TDC, achieving time synchronization across multiple modules. The measured time information is transmitted directly to the host computer via the Gigabit Ethernet port for real-time acquisition and processing.

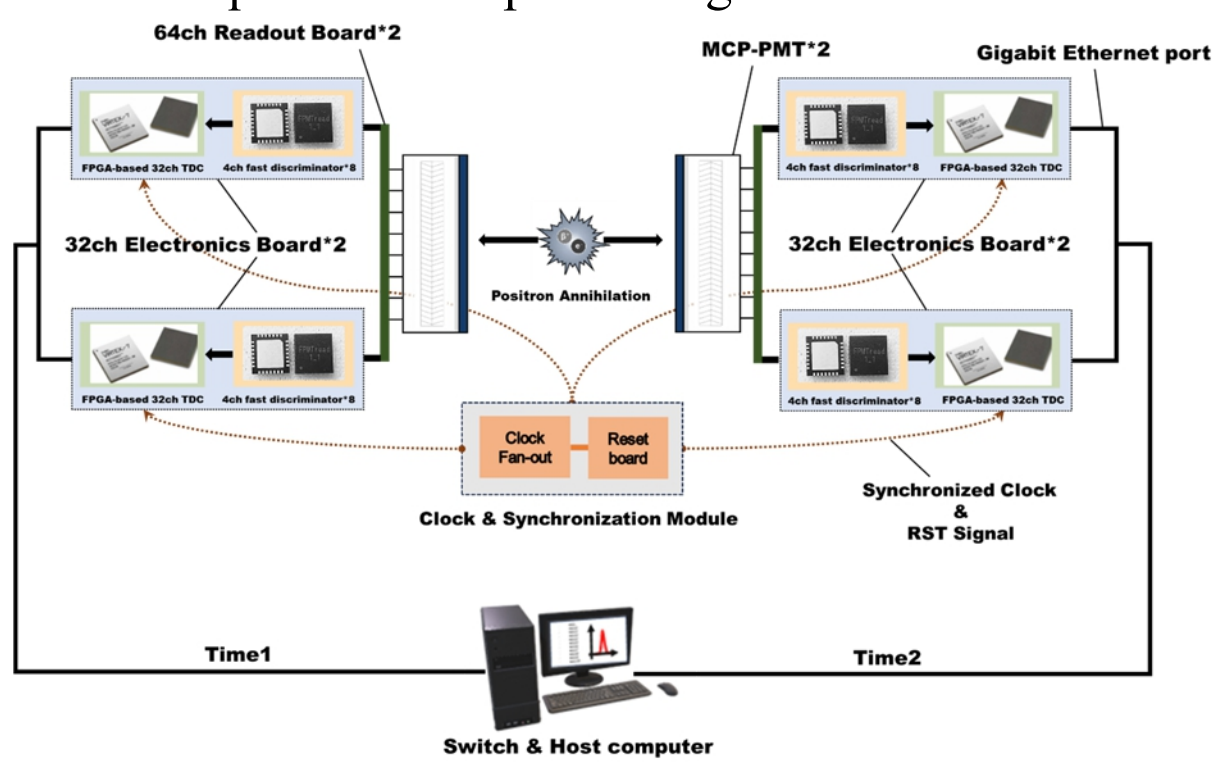

**Figure 4. Schematic diagram of the dual-module PET imaging platform structure and components.**

### *3.2 Preparation of Experiment*

After setting up the experiment, a series of calibrations and corrections are required to ensure consistency among the multiple channels of the detector module before conducting the experiment. As shown in Figure 5, the preparatory work for this experiment is as follows:

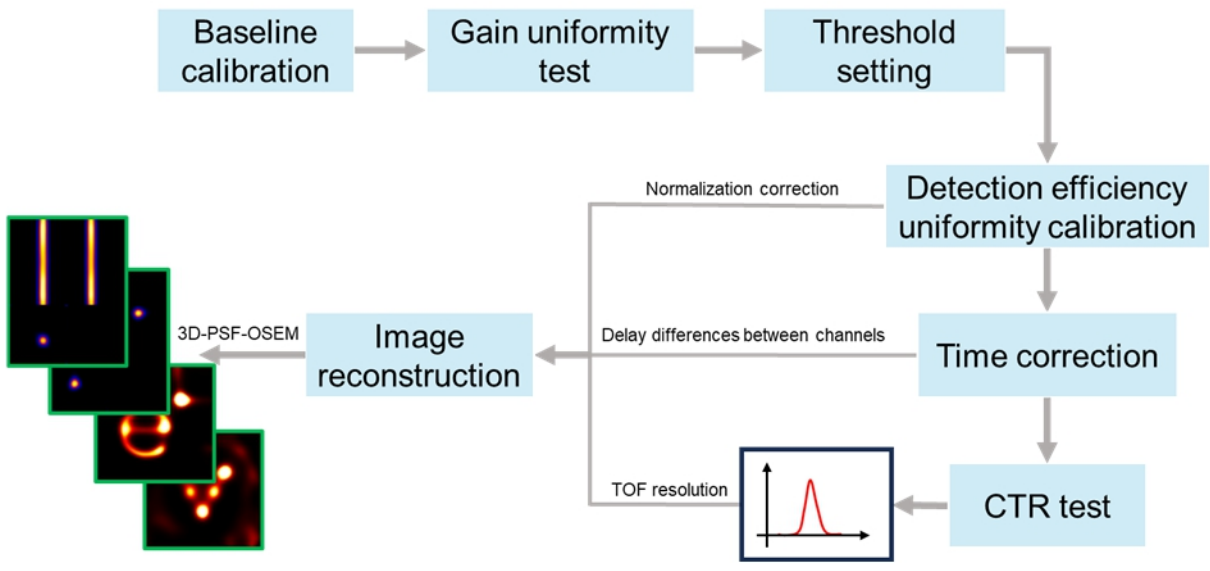

**Figure 5. Flowchart for calibration and correction of the dual-module imaging platform.**

### *3.2.1 Baseline Calibration and Threshold Setting*

As described above, in the working principle of the FPMTread discriminator ASIC chip, the input signal of the discriminator is added to a baseline. The threshold value needs to be set with reference to this baseline. However, due to process variations in chip manufacturing, the baseline of each channel varies. To accurately set the threshold value, it is necessary to calibrate the baseline of each channel individually. As shown in Figure 6a, we performed a threshold scan with an accuracy of 1 mV and recorded the value when the noise triggered as the channel baseline to obtain channel-by-channel baseline calibration data. The results are shown in Figure 6b and are used as references for subsequent threshold setting.

Referring to the previous results of the performance test of the sample Cherenkov radiator-integrated window MCP-PMTs (Ma *et al* 2023), the gain of different anodes of the Cherenkov radiator-integrated window MCP-PMTs varies and shows a geometrical pattern, with high gain in the center and low gain at the edge. The difference between the edge and the center is approximately 1.7 times. Based on this characteristic, we set the threshold "by ring", as shown in Figure 6c. The single-photon amplitude distribution of the center anode of the MCP-PMTs has a peak level of about 52 mV. Based on the timing threshold of a quarter of the single-photon amplitude, we set the thresholds from the center to the edge at 13 mV, 11 mV, 9 mV, and 7 mV, respectively.

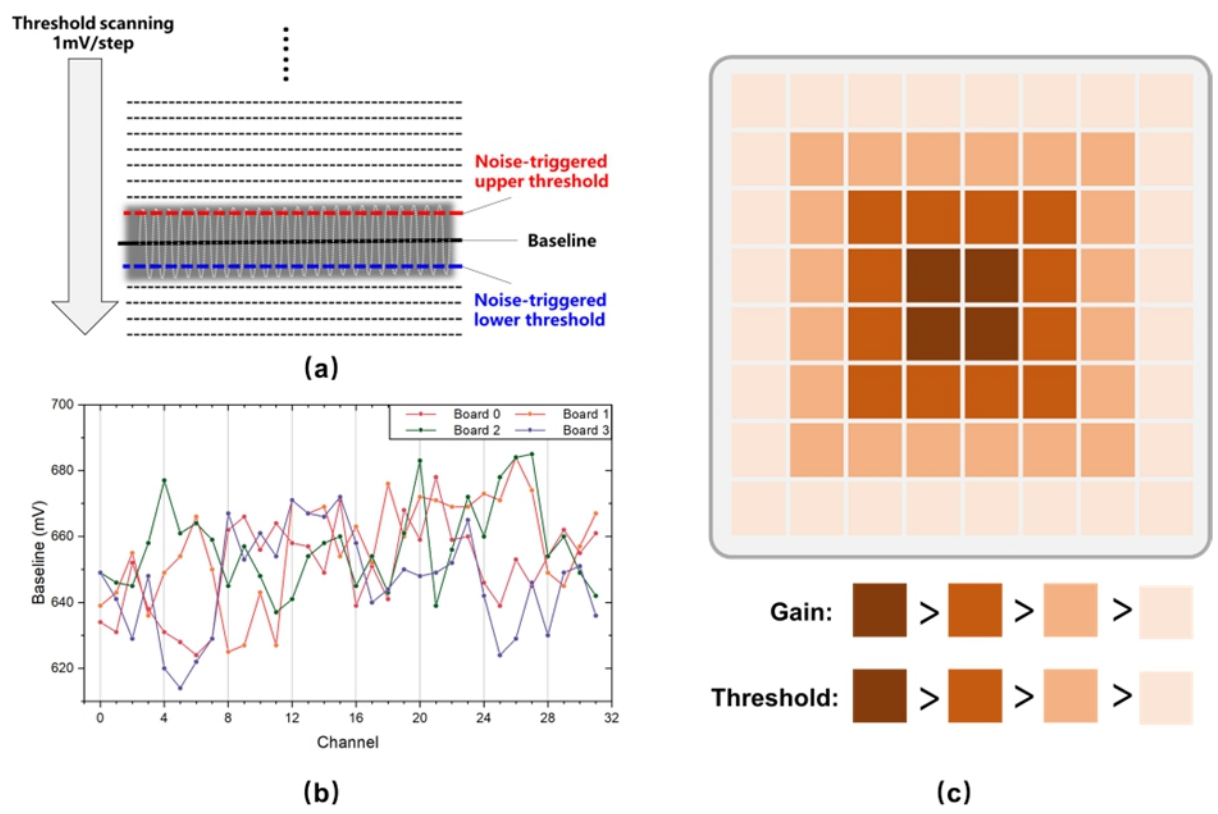

**Figure 6. Schematic diagram of the baseline calibration method(a) Baseline calibration results (b) and a schematic diagram of the "by ring" threshold-setting method (c).**

### *3.2.2 Detection Efficiency Uniformity Calibration*

To establish normalization correction parameters for the CASToR reconstruction and to verify the validity of the threshold-setting method and baseline calibration, we performed uniformity calibrations of the detection efficiency both before and after the baseline calibration and the "by ring" threshold setting. We used a $^{137}$Cs point source, as shown in Figure 7a, positioned approximately 300 mm from the detector to minimize counting differences between anodes due to solid angle variations. After placing the radioactive source, we evaluated the count rates of the different anodes and subtracted the pre-measured dark count rate to determine changes in the detector count rate response, which served as a reference for uniformity calibration. As shown in Figure 7b and Figure 7c, after normalizing the data, the standard deviation of the uniformity calibration results for the same detector decreased from 0.323 before baseline calibration and the "by ring" threshold-setting method to 0.190 afterwards. This indicates that the uniformity between detector channels was significantly improved after the channel-by-channel baseline calibration and the "by ring" threshold-setting method. Based on these optimized results, we obtained normalization correction parameters for each LOR, which were used for subsequent image reconstruction.

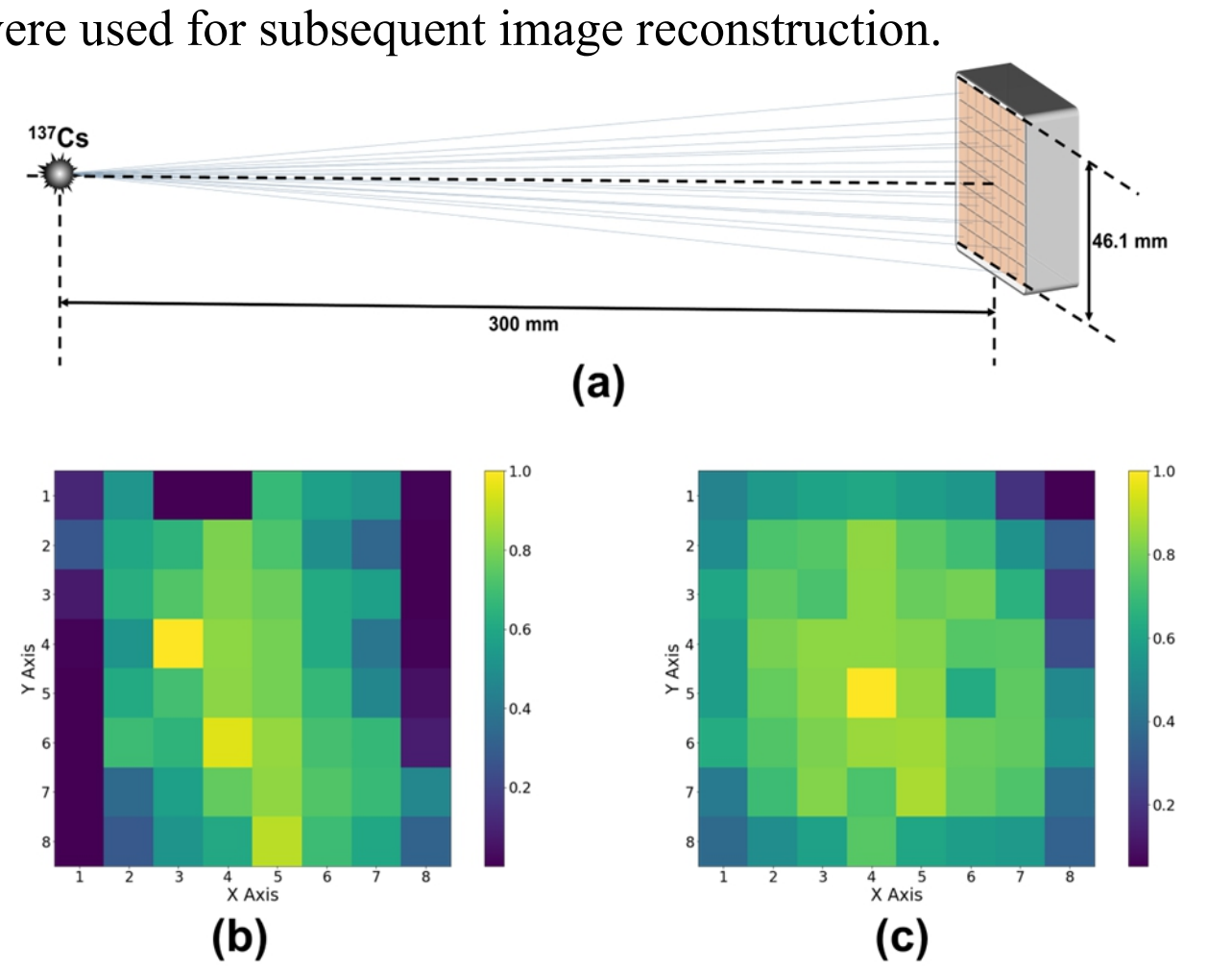

**Figure 7. Schematic diagram of the experimental setup for the detection efficiency uniformity calibration (a). Results of detection efficiency uniformity before (b) and after (c) the baseline calibration and the application of the "by ring" threshold-setting method.**

### *3.2.3 Time Correction*

In multi-channel time measurement experiments like this study, some fixed time delays between channels are unavoidable due to differences in signaling distances and the length of the delay chain within the FPGA. To improve time resolution and mitigate the effects of these fixed time delays, channel-by-channel timing calibration is essential. As shown in Figure 8a, we placed two detectors 30 mm apart and positioned a $^{22}$Na point source adjacent to one of the detectors.

The reason for setting the detector spacing to 30mm is to obtain a larger stereo angle coverage ratio and improve the efficiency of time-corrected calibration. We utilized the nearest anode as the reference channel and the 64 channels of the other detector as the channels to be corrected. This setup allowed us to obtain the distribution of time differences of coincident events between each detector channel and the reference channel. We then determined the peak positions and calculated the theoretical peak positions for each time distribution based on the optical path differences of the gamma photon pairs. Corrections were calculated for each channel based on the differences between these two values.

We applied the aforementioned method to perform time corrections for each channel. Before and after the correction, we selected a fixed reference channel and recorded the coincidence events between it and the 64 anodes of the other detector. We measured the peak positions of the coincidence time distributions for the 64 LORs. The results are shown in Figure 8b. After the correction, the peak positions of the coincidence time distributions for all LORs are close to 0 ps. This aligns with the position of the point source in the experiment, which is exactly in center of the field of view, thus confirming the reliability of the time correction.

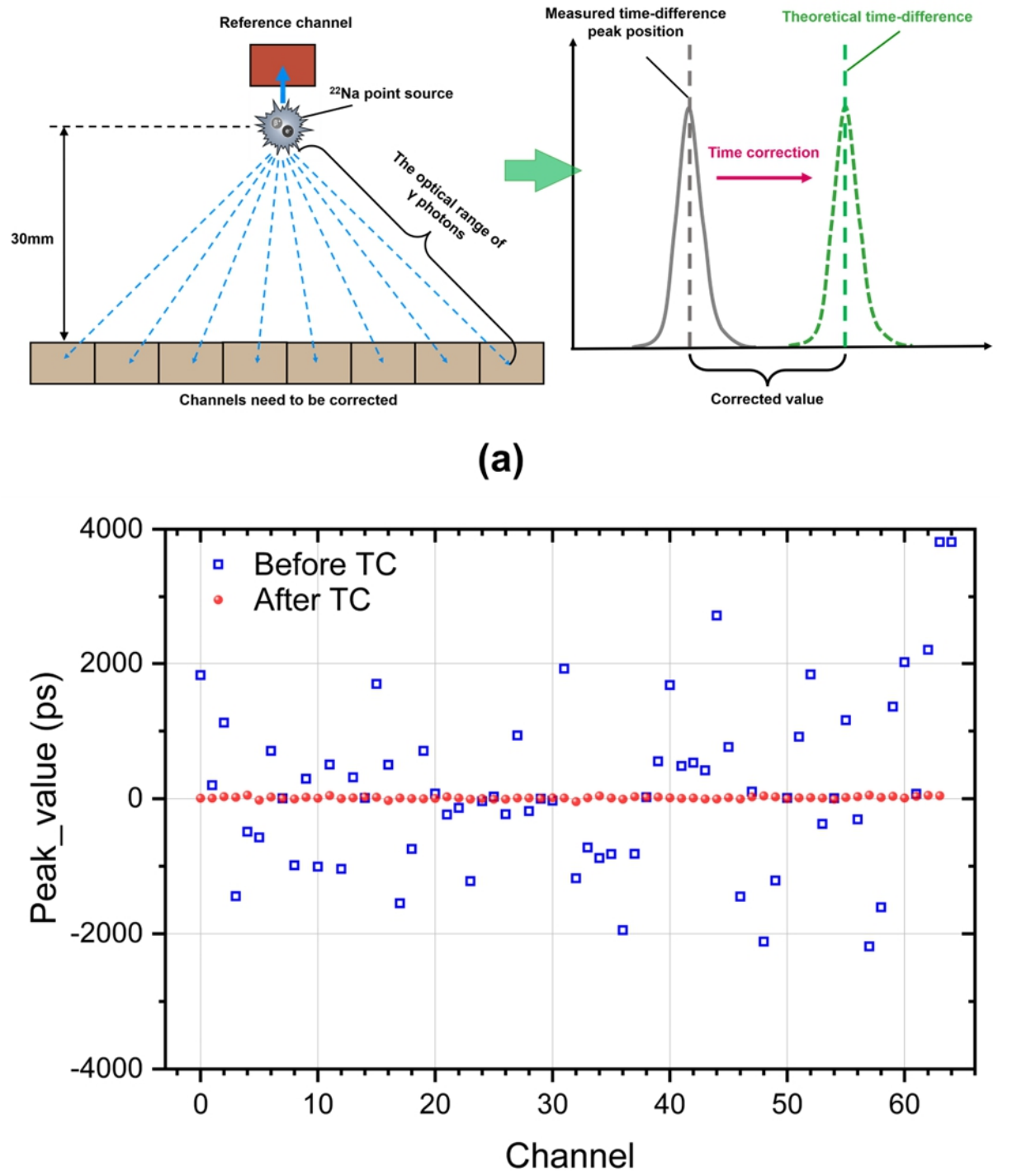

**Figure 8. Schematic diagram of time correction(a) and peak positions of time distributions before and after time correction on the sample LORs(b).**

### *3.3 Acquisition Method, Image Reconstruction and Correction*

As shown in Figure 9, the distance between two detector modules was set to be equal to the sensitive area of the detector (i.e., 46.1 mm), and the object to be imaged was placed at the center of the field of view and rotated once through 90° during the experiment to achieve comprehensive data acquisition from all four sides.

Image reconstruction and associated data corrections were performed using the CASToR software toolkit. The detectors were configured in CASToR according to the same setup described in the previous section on the acquisition method. The system matrix was generated, and the detection efficiency calibration results were organized on a line-by-line basis according to the requirements of the normalization correction file. After the CTR test, the TOF resolution factor is added to the reconstruction settings file according to the results, so as to limit the range of the reconstructed position of the point of annihilation when employing TOF information for assisted reconstruction.

In each CASToR-based PET image reconstruction in the following experiments, the listmode-based 3D-PSF-OSEM algorithm was used, and TOF information and normalization correction parameters were added to each set of data. Each reconstruction process was performed with 8 iterations containing 5 subsets, and the size of each voxel in the reconstruction was set to 0.5*0.5*0.5 $mm^3$.

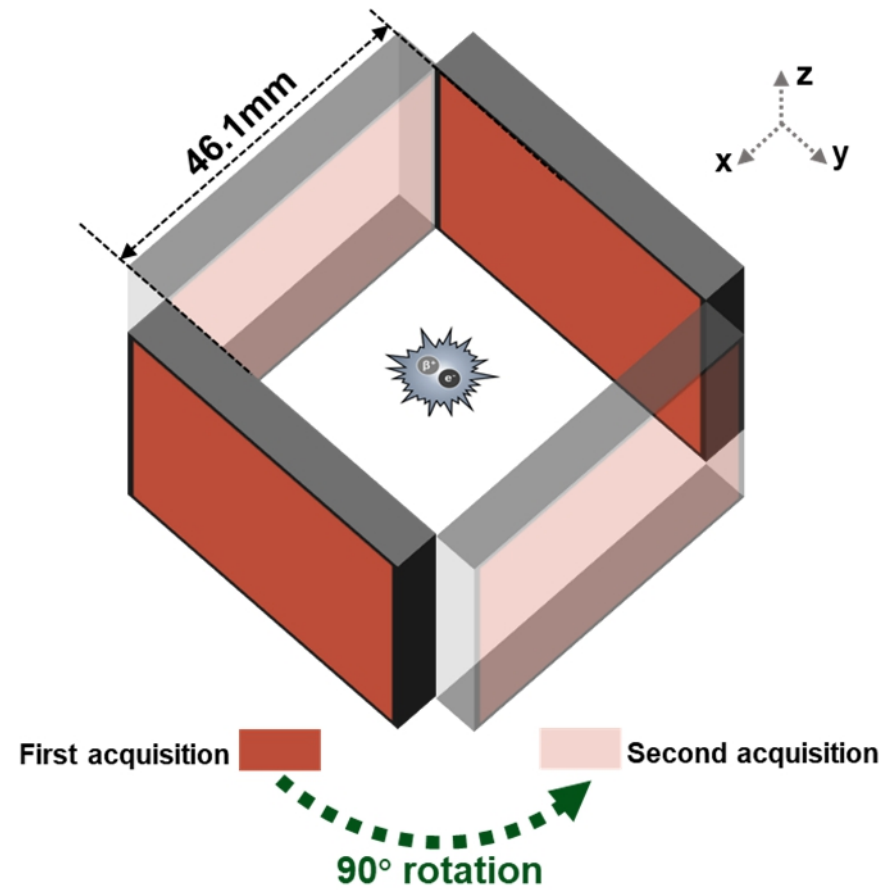

**Figure 9. Schematic diagram of the image acquisition method and the arrangement of detectors in CASToR**

## 4. Results

### *4.1 Point and Rod Source PET Imaging*

After all corrections and calibrations were completed, imaging experiments with point and rod sources were conducted according to the layout described in the previous acquisition method, as shown in Figure 9. The radioactive source in the point source experiment was $^{22}$Na, with a diameter of 0.25 mm and an activity of approximately 27 μCi, while the radioactive source in the rod source experiment was $^{68}$Ge, with a diameter of 2.36 mm, a length of 256 mm, and an effective activity of approximately 10 μCi. In the point source experiment, an acquisition time of 1 h was set for each angle. In the rod source experiment, an acquisition time of 2 h was set for each angle, taking into account the lower effective activity of the rod source.

Before the imaging experiment, the intrinsic resolution of the detector module was evaluated. Using an experimental setup consistent with the imaging experiment, the coincidence count of a pair of opposing anodes was recorded as the Na-22 point source was moved along the Z axis as shown in Figure 9 at 0.5 mm each step. The result is shown in Figure 10a. After fitting, the intrinsic resolution of the detector module was evaluated to be 5.35 mm FWHM.

As shown in Figure 10b, the system successfully achieved PET imaging based on Cherenkov light detection and reconstruction of a point source. In the imaging results, a straight line in the x-direction passing through the center of the point source image was plotted, and a Gaussian fit was applied to the distribution of pixel values along its path to evaluate the image resolution capability of the two-module system. The fitting result yielded a FWHM of 3.05 mm (1.30 mm σ).

At the same time, we also performed a simulation verification for point source imaging to compare the simulation and experimental results. In the simulation, we set up a detector layout with a "4-sided surround" structure that is consistent with the experimental setup. In the simulation, we only retained the coincidence data between the detector modules facing each other, which is more consistent with the setup in the experiment, because the experiment was completed by collecting data twice with 90° rotation. In the simulation, the $TTS_{@SPE}$ of the detector module was set to 70.65 ps FWHM (30 ps σ). The data obtained was reconstructed in a way that was completely consistent with the experimental data. The final image file obtained was analyzed using the same method as the experimental data, and the FWHM was 2.98 mm (1.27 mm σ). After comparison, there was not much difference between the results of the experiment and the simulation, and the two achieved mutual verification. These results provide preliminary verification of the feasibility of achieving PET imaging based on Cherenkov light detection using the integrated-window multi-anode MCP-PMT.

After completing the point source imaging experiment, a rod source imaging experiment was performed before and after using the normalization correction parameters, and the results are shown in Figure 10c. The reconstructed image without the normalization correction parameters is clearly missing at both ends. After the normalization correction parameters are added, the missing area is filled, improving the uniformity of the imaging results.

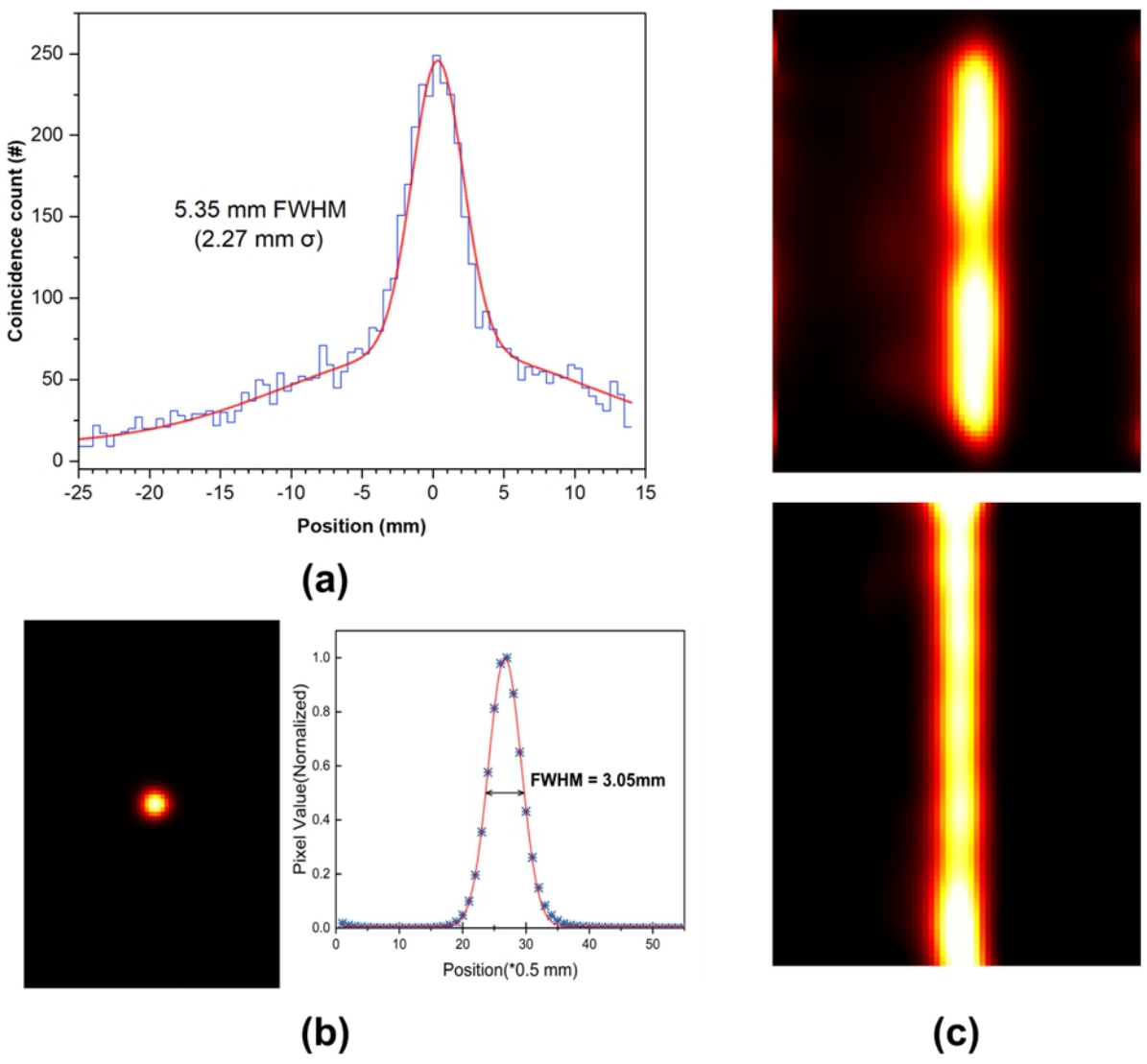

**Figure 10. Result of the intrinsic resolution test (a) and point source image results and analysis (b), rod source image before and after normalization correction (c).**

### *4.2 CTR Test*

The coincidence time resolution of the platform was analyzed based on data from the point source imaging experiments. All data on the lines of LORs with a certain level of coincidence counts were summarized to obtain an overall time distribution for the platform, as shown in Figure 11. The overall coincidence time resolution was determined to be 203.6 ps FWHM.

In this version of the integrated-window multi-anode MCP-PMT, although a solution has been identified during the process and will be implemented in subsequent prototypes, the microchannel plate (MCP) material currently used still contains a certain amount of lead. Therefore, the microchannel plate and the annihilation gamma rays also have a certain cross section of interaction, based on the coincidence events in which the two gamma rays interact with the MCP-PMT. Specifically, these are categorized as follows: radiator optical window-radiator optical window, MCP (left)-radiator optical window, radiator optical window-MCP (right), and others (includes photo-electron backscatter and MCP-MCP interactions, etc.) (Ota *et al* 2020a). The coincidence events are divided into four categories, and a 4-Gaussian fit to the total time distribution was performed (see Figure 11), with the specific parameters of each Gaussian peak shown in Table 5.

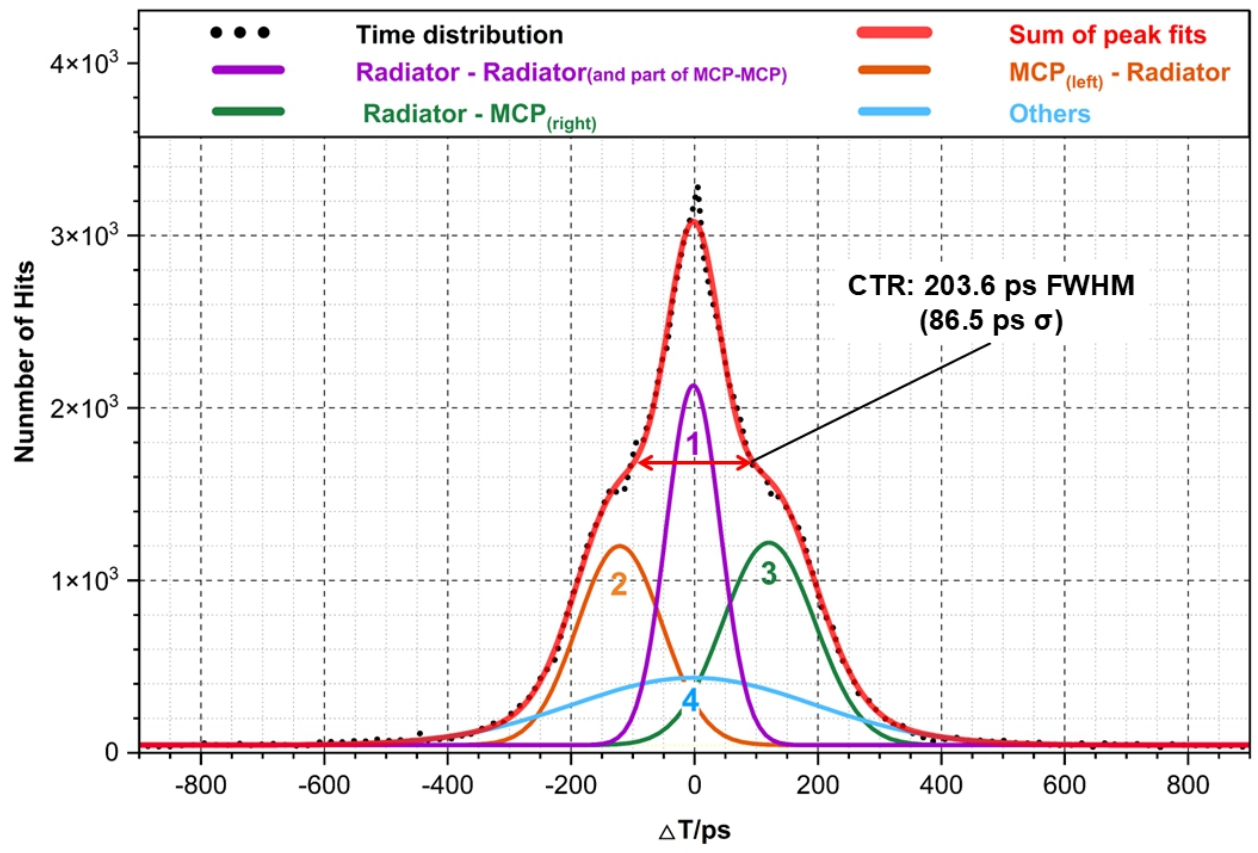

**Figure 11. The overall time distribution and the time distribution decomposed by four categories of coincidence event: radiator optical window-radiator optical window and part of MCP-MCP (1), MCP (left)-radiator optical window (2), radiator optical window-MCP (right) (3), and others (includes photo-electron backscatter and MCP-MCP interactions, etc.) (4).**

**Table 5. Fitting parameters for a four-Gaussian fit.**

| Peak number | Type of interaction | Xc | FWHM | Event ratio |
|---|---|---|---|---|
| 1 | Radiator – Radiator (and part of MCP-MCP interactions) | -1.97 ps | 103.17 ps (43.81 ps σ) | 27.32% |
| 2 | MCP (left) - Radiator | -121.19 ps | 161.60 ps (68.62 ps σ) | 22.75% |
| 3 | Radiator – MCP (right) | 120.46 ps | 176.12 ps (74.79 ps σ) | 26.25% |
| 4 | Photo-electron backscatter and MCP-MCP interactions, etc. | -1.96 ps | 458.79 ps (194.82 ps σ) | 22.68% |

In Figure 11 and Table 5, the four peaks numbered 1 to 4 correspond to the coincidence events under the four interaction modes described above. The estimation of the time of drift of the photoelectron in the electric field between the cathode and the MCP was carried out according to the calculation process shown in equations (1) to (3). First, the method for calculating the electric field strength between the photocathode and the MCP is given in equation (1), Here $U$ is the voltage difference between the photocathode and the MCP, which is 130 V, and $d$ is the geometric spacing between the photocathode and the MCP, which is 0.4 mm, as shown in Figure 1b, from which the electric field strength $E$ between the photocathode and the MCP is obtained.

$$E = \frac{U}{d} \tag{1}$$

Subsequently, the method for calculating the acceleration of an electron in an electric field is shown in Equation (2), where $a$ is the acceleration of an electron in an electric field, e is the elementary charge, and m is the mass of an electron.

$$a = \frac{e \times E}{m_e} \tag{2}$$

Finally, the drift time is calculated. By bringing all the values into equation (3), the drift time of the electron is obtained to be approximately 118 ps.

$$Drift\ time = \sqrt{\frac{2 \times m_e \times d^2}{e \times U}} = 118\ ps \tag{3}$$

The peak differences between the two side peaks and the main peak are 122 ps and 119 ps, respectively, which corresponds to the electron drift time. This further confirms that the cause of the side peaks is indeed the annihilation gamma rays interacting with the MCP material. This problem will be addressed by eliminating the lead in the MCP material and further optimizing the blocking capability and detection efficiency of the optical window material. The main peak numbered 1 has a 103 ps FWHM. This represents the level of time resolution obtained without any data filtering or correction. This serves as an initial demonstration of the advantages and potential of PET imaging based on Cherenkov light detection and the integrated-window MCP-PMT in improving time resolution.

### *4.3 e+ Phantom Imaging*

In addition, we demonstrated the PET imaging results of the platform for structured objects. Considering the size of the dual module and the experimental setup, we specifically designed the $e^+$ shaped phantom. The design of the phantom is shown in Figure 12a, with a side length of 35 mm and a tube diameter of 1.5 mm. Before the phantom imaging experiment, ~200 μCi of $^{18}$F ionic solution was injected into the phantom. Due to the short half-life of F-18, in order to avoid the difference in the coincidence counting between different acquisition angles caused by the decay process, the imaging experiment of the phantom was reconstructed using the data from a single angle, and the acquisition time was 1800 s.

The imaging result of the $e^+$ module is shown in Figure 12b. To the best of our knowledge, this is the first study to achieve module-level PET imaging based on Cherenkov light detection. This verifies the feasibility of Cherenkov imaging at a certain scale. Although there is still some degree of non-uniformity between the channels, resulting in highlights at the corners of the phantom and unclear representation at the "+" region, overall, it clearly reflects the majority of the structural information of the $e^+$ phantom, and these issues will be addressed in subsequent efforts to optimize homogeneity and apply better normalization corrections.

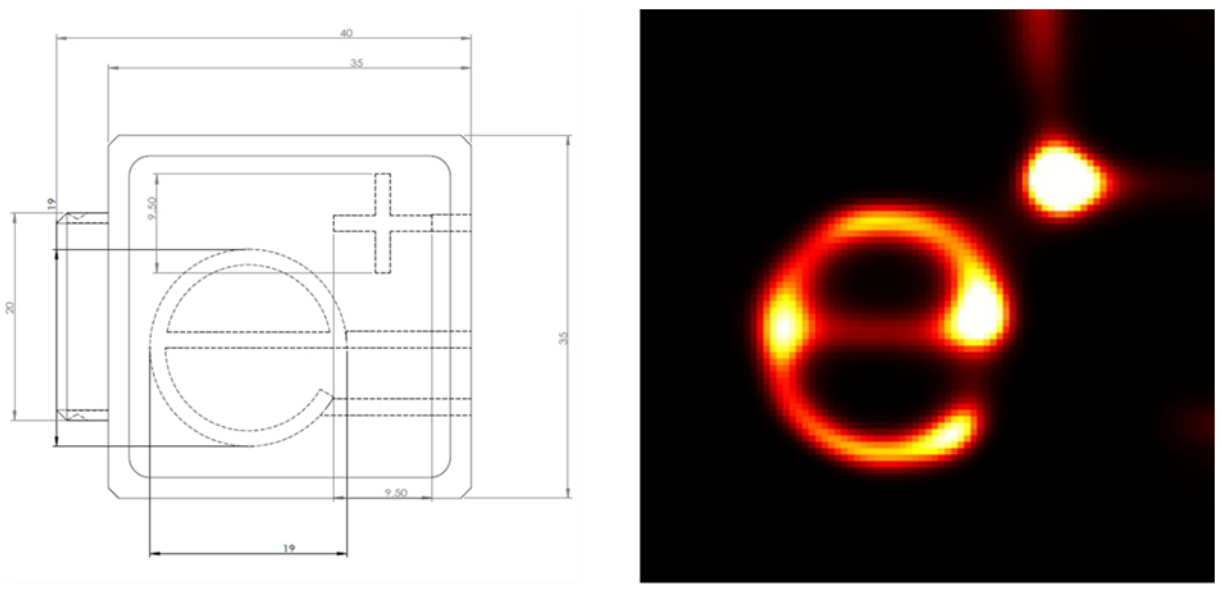

(a) (b)

**Figure 12. Phantom design (a) and Experimental result of e+ phantom imaging (c)**

### *4.4 Derenzo Phantom Imaging Experiment and Spatial Resolution*

After verifying that PET imaging based on the detection of Cherenkov light can image structural objects, a PET imaging experiment of the Derenzo phantom was performed to evaluate the spatial resolution of the platform. Before performing PET imaging experiments based on the Derenzo phantom, the theoretical spatial resolution of the dual-module imaging platform was analyzed. The theoretical relationship between the spatial resolution Γ of the PET system and the corresponding parameters is shown in equation (4), where $d$ is the pixel width, $s$ is the positron range, and $D$ is the detector ring diameter(Moses 2011).

$$\Gamma = 1.25\sqrt{\left(\frac{d}{2}\right)^2 + s^2 + (0.0022D)^2} \quad (4)$$

According to the platform parameters, all parameters were brought into the above theoretical analysis formula, where the width of the detector pixel is 5.5 mm, the range of the positron is 0.54 mm due to the use of the positron nuclide $^{18}$F, and the diameter of the detector ring is 46.1 mm. The final theoretical analysis obtained the spatial resolution of the platform as shown in equation (5), and the final spatial resolution of the dual-module imaging platform obtained based on theoretical analysis is about 3.51 mm FWHM (1.49 mm σ).

$$\Gamma = 1.25\sqrt{\left(\frac{5.5}{2}\right)^2 + 0.54^2 + (0.0022 \times 46.1)^2} \approx 3.51\ mm\ FWHM \quad (5)$$

Using the spatial resolution obtained from theoretical analysis as a reference, an imaging experiment was carried out to verify the Derenzo phantom. According to the analysis of the spatial resolution, the Derenzo phantom model most suitable for the dual-module conformal imaging platform is the Mini Hot Spot Phantom(Spectrum 2025), which has hollow channels with diameters of 1.2, 1.6, 2.4, 3.2, 4, and 4.8 mm. However, the effective area diameter of the aforementioned Derenzo phantom is 75 mm, and the effective field of view of the dual-module platform is only 46.1 mm, so it is difficult to meet the acquisition requirements of the above-mentioned phantom. Therefore, we have designed three phantoms based on the same hollow channel distribution law, with hollow channel diameters of 2 mm, 3 mm, and 4 mm, respectively. The phantom design is shown in Figure 13a and is consistent with the distribution rule of the hollow channels in the standard Derenzo phantom, that is, the distance between the centers of adjacent hollow channels is twice the diameter of the hollow channel.

Imaging experiments were performed based on three phantoms, respectively. Consistent with the imaging experiment of the "e$^+$" phantom, a single-angle acquisition was used, and a solution of F-18 ions with an activity of ~200 μCi was injected. Each phantom was scanned for 1800 s. The summary image results of the three phantoms are shown in Figure 13b.

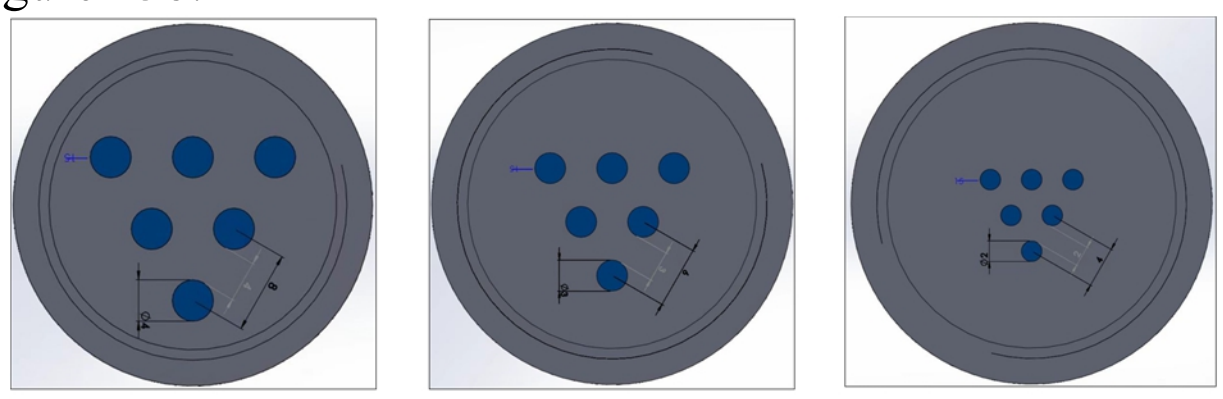

(a)

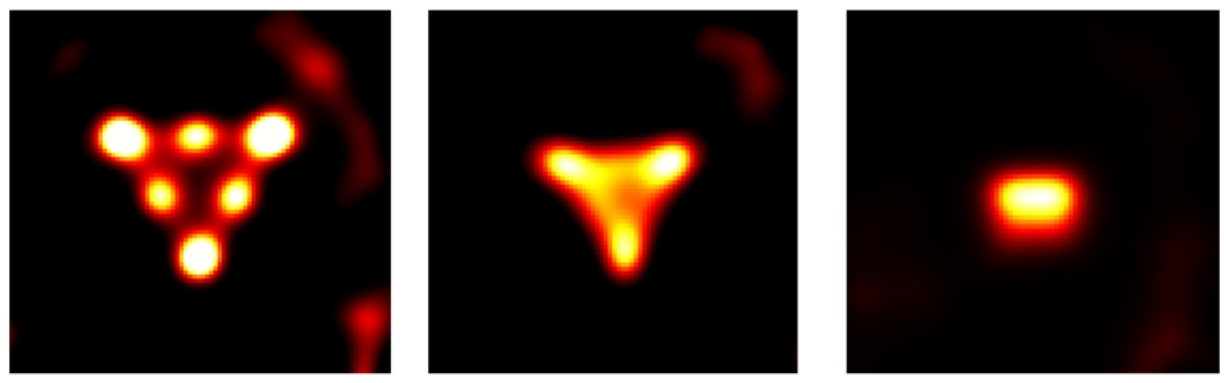

(b)

**Figure 13. Custom-designed Derenzo phantom design (a) and corresponding imaging results (b)**

The imaging experiment results for the Derenzo phantom show that although there are certain problems with the uniformity of the image due to the lack of uniformity in the detector channels, the 4 mm diameter rods are well distinguished, while the 3 mm and 2 mm rods are more difficult to distinguish. This is consistent with the theoretical analysis of the spatial resolution of the two-module platform described above.

Based on experimental data from 4 mm Derenzo-like phantoms, we performed image reconstruction with and without TOF information, as shown in Figure 14. We plotted profile curves for three points corresponding to the top of the phantom in the images and analyzed the valley-to-peak ratio (VPR). The minimum valley-to-peak ratio in the reconstruction image without TOF information was 0.242, while the corresponding result in the image with TOF information was 0.117. The results indicate that, based on the experimental setup described in this paper, the current level of time resolution can already bring about preliminary improvements in image quality.

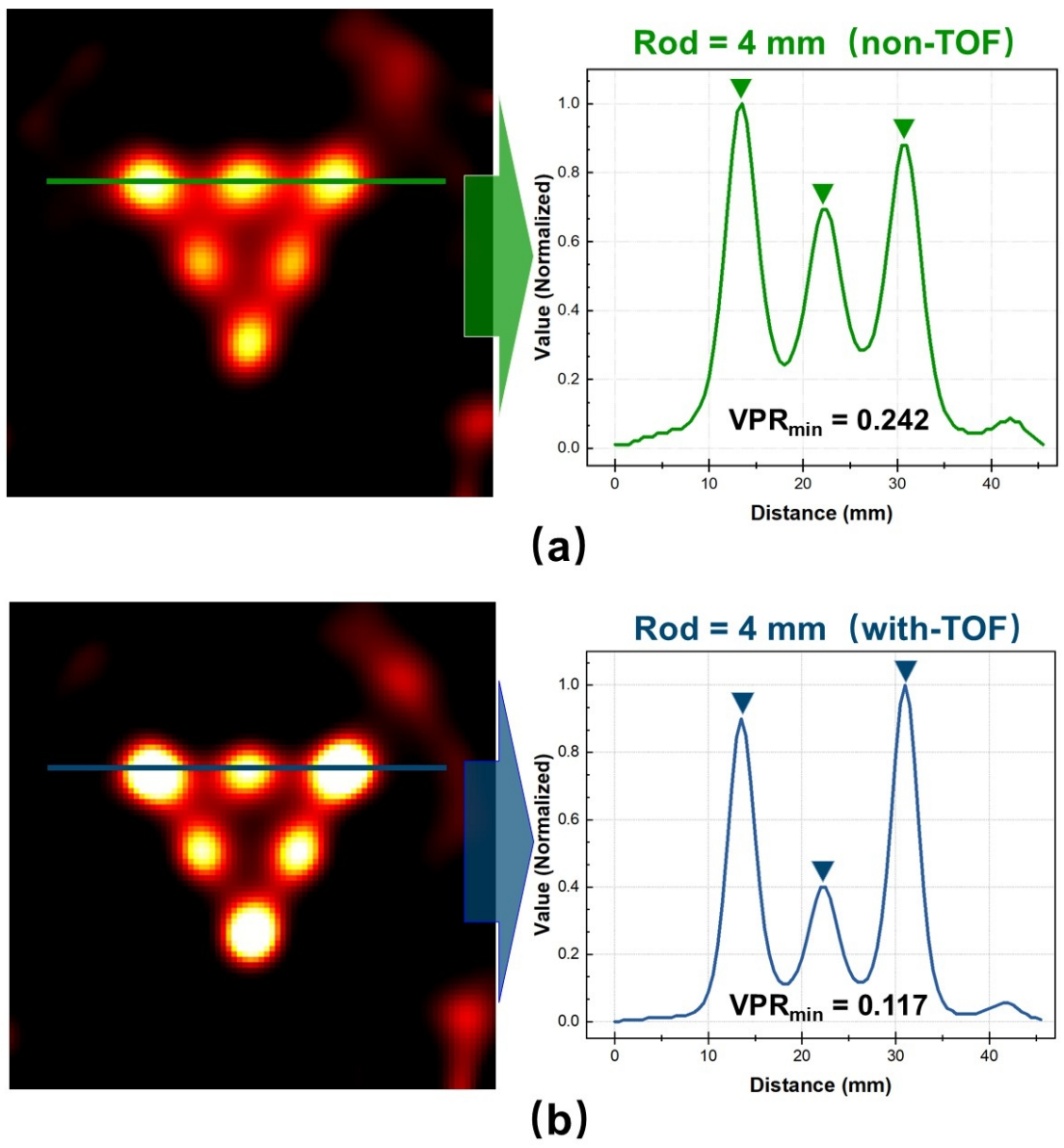

**Figure 14. Image results of 4 mm Derenzo-like phantoms without (a) and with (b) TOF information, along with their profile curves and corresponding valley-to-peak ratio analysis.**

## 5. Discussion

In this study, we developed a dual-module PET imaging platform utilizing a multi-anode MCP-PMT with an integrated window of Cherenkov radiators, performed specialized calibrations and corrections to account for the unique characteristics of the platform, evaluated its timing performance, and conducted preliminary imaging experiments.

### *5.1 Comparison of this study with previous relevant studies*

#### *5.1.1 Improved system feasibility and imaging efficiency brought by multi-anode MCP-PMT*

Prior to this study, PET imaging studies based on Cherenkov light detection had already been conducted using single-pixel devices. In recent research, an excellent time resolution of 26.4 ps FWHM was achieved, and dPEI was accomplished using a single-anode MCP-PMT with integrated window material (Kwon *et al* 2021b). However, because the single-anode MCP-PMT lacks positional resolution capability, the localization of the radiation source is accomplished through collimation and translation of the detector, which significantly impacts imaging efficiency. This represents the most significant obstacle to the practical implementation of this modality in PET.

Our self-developed multi-anode MCP-PMT not only increases the sensitive area and reduces the proportion of dead space, but its square structure also meets the needs of multi-module assembly. Most importantly, the multi-anode design endows the detector with position resolution capability, so that collimation and translation are no longer required during imaging. In previous studies based on single-anode MCP-PMT, only the coincident events on one LOR line could be obtained per acquisition. However, after the multi-anode MCP-PMT design is implemented, the number of LOR lines with coincidence events per acquisition will be the square of the number of anodes. The self-developed multi-anode MCP-PMT has an anode arrangement of 8 × 8, so the number of LOR lines that can be covered in one acquisition in this study will be 4096 times that of the study based on a single anode MCP-PMT. Therefore, although the time resolution achieved by the dual-module imaging platform we built (203 ps FWHM) is still somewhat inferior to the above-mentioned study, the imaging efficiency has been significantly improved compared to the study based on the single-anode MCP-PMT. Unlike the imaging experiments based on the single-anode MCP-PMT, which often require the injection of more than 2 mCi of positron nuclides and a collection time of 4-24 h. In this study, the dual-module imaging platform based on the multi-anode MCP-PMT only requires the injection of ~200 μCi of positron nuclides and ~30 min of acquisition time for a complete image to be obtained for a similar phantom, which reflects the significant improvement in imaging efficiency brought by multi-anode MCP-PMT.

Moreover, by utilizing self-developed multi-channel high-precision electronics instead of a high-precision oscilloscope, power consumption is reduced from more than 50 W/ch to 169 mW/ch (the power supply voltage and current of the self-developed 32-channel electronic device are 12 V and 451 mA, respectively). This also allows the time measurement-related electronics to be placed closer to the detector, which is beneficial for improving the accuracy of the time measurement. In addition, it overcomes the limitations of high-performance oscilloscopes in terms of the number of channels (usually <8 channels) and cost control (often >$1,000 per channel), thereby improving the efficiency of time measurement, data acquisition and processing, making it more feasible for practical applications.

#### *5.1.2 dPEI based on multi-anode MCP-PMT*

Since the current time resolution of the dual-module imaging platform based on the multi-anode MCP-PMT is only 203 ps FWHM, and this time resolution is only obtained by analysing the events in which the interaction occurs in the Cherenkov radiator window, there is still a long way to go before dPEI can be achieved based on the multi-anode MCP-PMT. However, in our previous research, by using the pulse-cutting analysis signal screening method (Ota *et al* 2021), we have verified in oscilloscope-based single-channel waveform tests that the current multi-anode MCP-PMT can achieve a time resolution of about 57.3 ps FWHM (24.3 ps σ) using this method (Chen *et al* 2024), but only about 6% of the coincidence events were retained. We refer to this method and use the pulse width information obtained by the dual-edge TDC measurement to perform pulse width-based signal screening in a dual-module imaging platform, and carry out time resolution verification after applying the pulse-cutting analysis signal screening method. Finally, after screening the top 40% of the pulse width distribution, about 9% of the coincident events were retained, and the resulting time spectrum is shown in Figure 15. The results show that after pulse width screening, the side peak events caused by the interaction of the ray and the MCP are significantly suppressed, and a time resolution of ~58.9 ps FWHM (25.0 ps σ) similar to that in the above study is achieved at the module level.

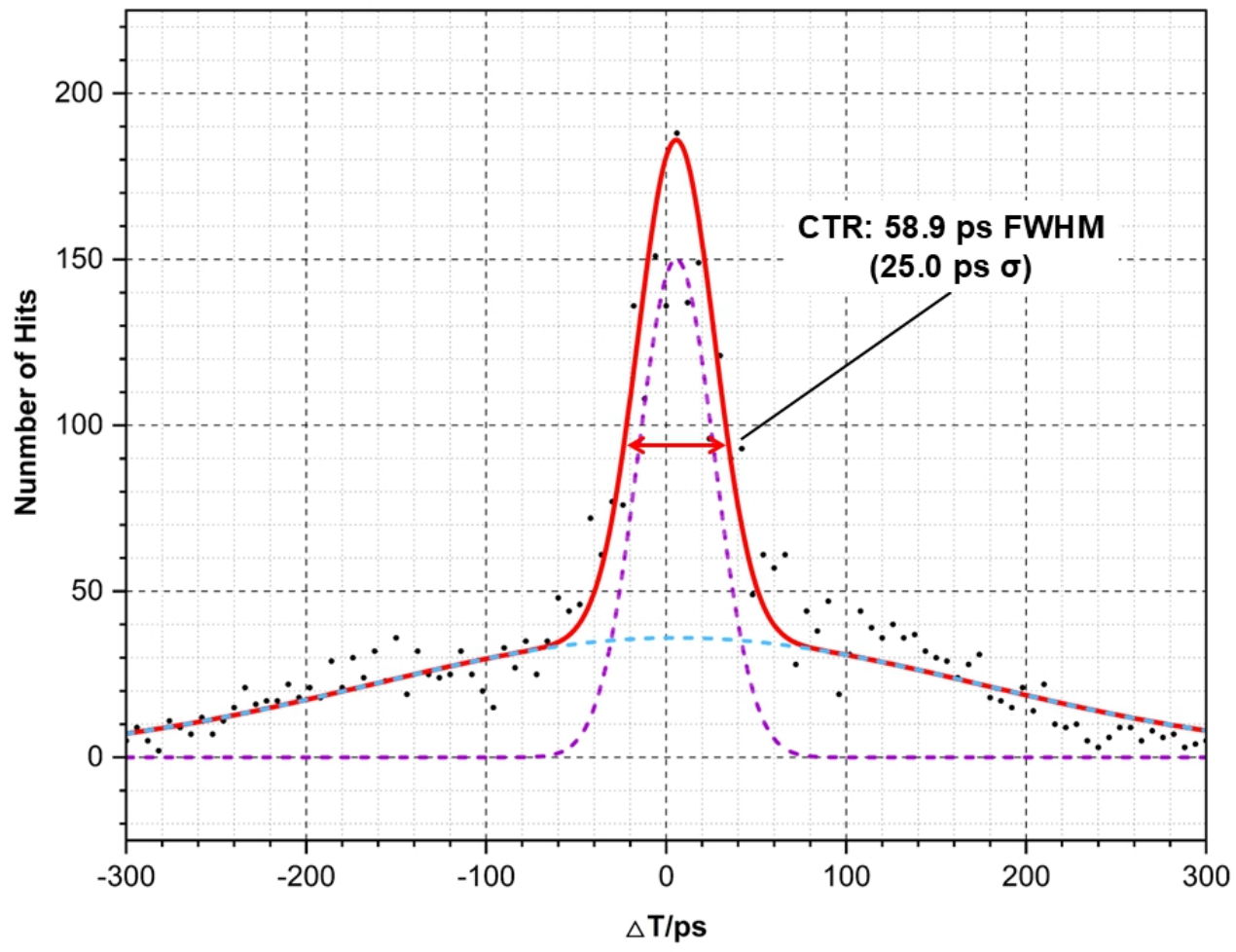

**Figure 15. Time resolution result at the module level after using a pulse width-based signal screening method.**

Based on these results, although most of the events were filtered out, the potential of the dual-module imaging platform based on multi-anode MCP-PMT for achieving higher time resolution and dPEI was verified. It also provides the basic conditions for preliminary verification of dPEI based on multi-anode MCP-PMT.

However, with the improvement of time resolution and the increase of the number of anodes, the scale and complexity of the correction and pre-calibration work, as well as the required precision, will increase significantly. Unlike dPEI based on single-anode MCP-PMT, which only requires high-precision time measurement of a single pair of anodes and a LOR line, dPEI based on multi-anode MCP-PMT needs to complete time calibration and performance calibration between the number of LOR lines and channels that are geometrically multiplied, and maintain high time measurement precision. In addition, the loss of event rate due to pulse screening makes it challenging to implement dPEI at the module level. How to achieve efficient performance calibration and time correction without losing measurement precision will be the key, which will also be the focus of our upcoming research. At the same time, continuously improving the detection efficiency and time performance of the self-developed multi-anode MCP-PMT, and optimizing its uniformity between channels, is also a priority.

### *5.2 The importance of further improving the consistency between different anodes in MCP-PMT*

It is important to note that, although PET imaging based on Cherenkov light detection at the module level has been achieved, the uniformity of the system in current experiments remains suboptimal. This is largely attributed to the unique characteristics of the Cherenkov mechanism. Due to the energy of the annihilated gamma rays, the detection of Cherenkov light often occurs at the level of a few photons or even at the single-photon level. Consequently, the gamma-ray signal cannot be distinguished from the dark noise of the detector itself. If the dark noise level is high or unstable, it will directly impact the calibration and correction results of the detector. Additionally, the relatively weak signal exacerbates the effects of gain variations among detector pixels and response differences between electronic channels. As noted earlier, the uniformity of the system has been initially addressed through complex calibration and specialized threshold-setting methods; however, normalization correction remains necessary, and the quality of the image is highly dependent on the reliability of this normalization correction. Therefore, subsequent work will also focus on enhancing the consistency of the detector itself, while simultaneously exploring more reliable calibration and normalization correction methods to minimize the impact of non-uniformity on imaging.

### *5.3 The effect of the interaction between the annihilation photon and the MCP*

In the time distribution results, there are annoying side peaks symmetrically located on both sides of the main peak. After analyzing the experimental results and referencing previous studies (Ota *et al* 2019b, 2020b), it was determined that the source of the side peaks is the interaction between the annihilation gamma rays and the MCP. The presence of such peaks not only affects the overall time resolution and efficiency, as seen in single-anode applications, but also directly affects the spatial resolution in position-sensitive detector applications. As shown in Figure 16, this directly shifts the positioning of the annihilation point. Therefore, it is crucial to suppress this aspect of the events as much as possible. Reducing or even eliminating the lead content in the MCP is the most direct approach to suppressing the interaction between radiation and the MCP. The efficacy of this approach has been confirmed in previous studies, and we are currently optimizing it for this purpose, which will be implemented in the next version of the Cherenkov radiator-integrated window MCP-PMT.

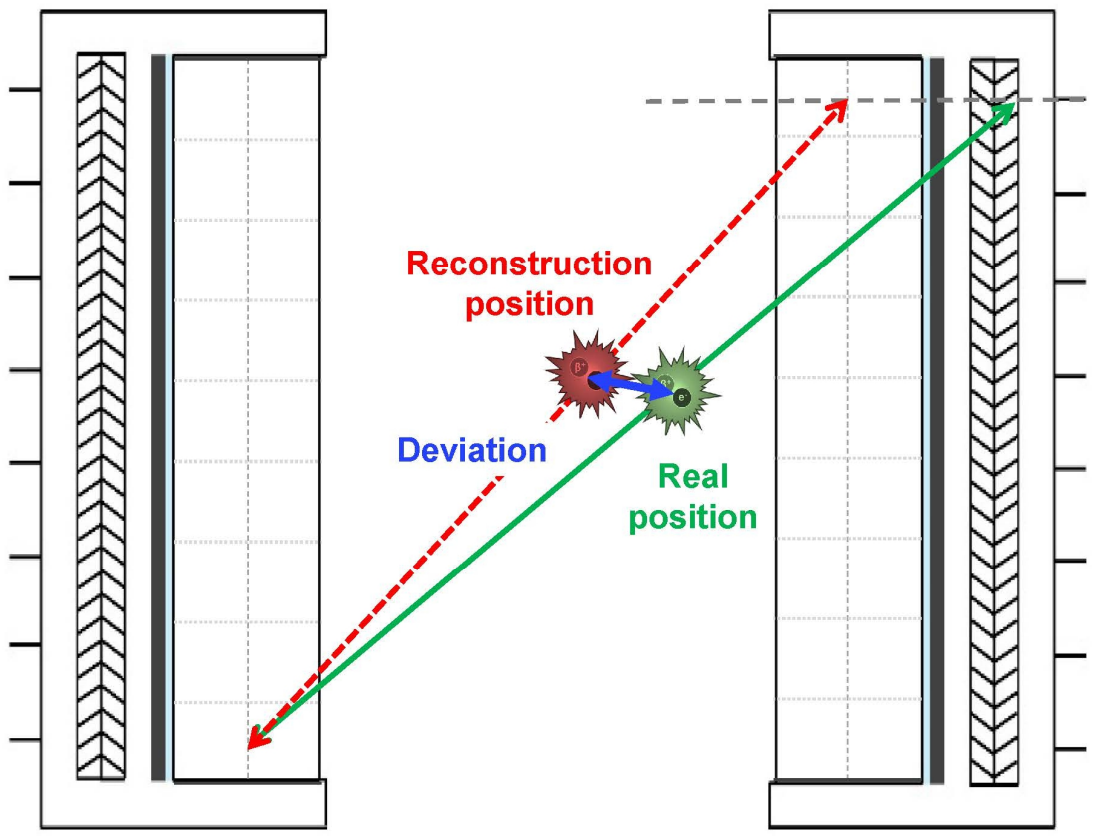

**Figure 16. Illustration of the positioning error caused by the interaction between the annihilation gamma rays and the MCP.**

In addition, the lead glass currently used is not the best material for Cherenkov radiator-integrated window. The reason for using lead glass as the window material is that it is less challenging to process to achieve an integrated window. Therefore, the development and proof of principle and feasibility of the integrated window multi-anode MCP-PMT was first completed based on lead glass. In fact, taking into account the core factors such as density, effective atomic number and cut-off wavelength, $PbF_2$ is actually a better material for Cherenkov integrated window. A comparison of the relevant parameters with the scintillation crystal LSO used in the current mainstream PET system is shown in Table 6.

**Table 6. Comparison of the parameters related to stopping power and detection efficiency between $PbF_2$ and lead glass and LSO**(Kratochwil *et al* 2021, Lewellen 2008)**.**

| | $PbF_2$ | Lead glass | LSO |
|---|---|---|---|
| Type of material | Cherenkov radiator | | scintillator |
| Density g/mm$^3$ | 7.77 | 5.49 | 7.4 |
| Refractive index n | 1.78-1.94 | 1.81-1.95 | |
| Effective atomic number $Z_{eff}$ | 77 | 62.5 | 66 |
| Cut-off wavelength nm | 260 | 370 | |

By optimizing the Cherenkov integrated window material, the number of photoelectrons ultimately output by the interaction of gamma rays and the window material will also increase. According to the law that the TTS of the detector is inversely proportional to the square of the number of photoelectrons $n^2$, the time resolution will be significantly improved. In addition, the new scintillator material has a higher stopping power, which further relatively suppresses the effect of gamma rays interacting with the MCP and reduces its impact on time resolution and annihilation point localization.

### *5.4 Influence of PET detection using Cherenkov light on the performance of PET system*

One drawback of PET detectors based on pure Cherenkov radiation is that, due to the low Cherenkov light yield corresponding to the 511 keV energy of annihilation photons, such detectors are difficult to achieve good energy resolution, which is often the reason why this PET detection method is questioned as to whether it can really improve the performance of the PET system. However, through the analysis of the principle of the Cherenkov radiation process, it is found that charged particles need to exceed the speed of light in the medium to generate Cherenkov light. Therefore, there is a threshold electron kinetic energy related to the refractive index of the medium material to determine whether Cherenkov radiation can be achieved. As shown in Table 1, taking the lead glass material used in this study as an example, this threshold electron kinetic energy is at least 85.3 KeV. The existence of this threshold kinetic energy to a certain extent naturally screens some scattering events and reduces the interference of scattering events on PET imaging. This is also the core role played by the energy resolution of the PET detector based on the traditional scheme. However, despite the existence of this threshold kinetic energy, it can only screen out lower-energy scattered events, and it will still cause the Cherenkov light-based PET detector to have a higher proportion of scattered events compared to the traditional PET detector solution. This requires higher detection efficiency and time resolution to compensate for the interference caused by scattered events.

As mentioned in the previous section, lead glass is not the best material for a Cherenkov radiator window. Its disadvantages in terms of cutoff wavelength, material density and effective atomic number mean that the detection efficiency achieved by using lead glass as an integrated window material for Cherenkov radiation is unsatisfactory. Even with excellent time resolution, it cannot effectively overcome the adverse effects of an increase in scattering events, and improve the overall performance and image quality of PET system. The experimental process and results of this study also verify this point of view. In fact, the 103ps FWHM time resolution achieved by the dual-module imaging platform we have built currently already shows certain advantages over the mainstream PET systems, but the imaging efficiency and image quality obtained are still far from those of PET systems based on scintillators.

However, when the Cherenkov radiator material is replaced with $PbF_2$, the results will be completely different. As shown in Table 6, in terms of parameters related to stopping ability and detection efficiency, $PbF_2$ even has a significant advantage over the current mainstream scintillator LSO, which makes PET detection based on $PbF_2$ as a Cherenkov radiator material have the advantages of both detection efficiency and time performance In fact, relevant studies have been conducted on the systematic simulation of PET systems based on $PbF_2$ as a Cherenkov radiator material(Consuegra *et al* 2020), The simulation results indicate that images obtained from a Cherenkov whole-body TOF-PET scanner using $PbF_2$ exhibit image quality metrics comparable to those obtained from a whole-body LSO TOF-PET scanner. It is worth noting that the photodetectors used in the aforementioned simulations employed typical performance parameters of commercially available SiPMs. If the photodetectors are optimized to the Cherenkov radiation-integrated window MCP-PMT with lower dark count rate and better time performance as described in this paper, the corresponding image quality metrics would further improve, potentially enabling performance surpassing that of LSO-based TOF-PET scanners.

Based on the scheme of this study, the improvement in PET system performance brought about by the use of a Cherenkov radiator integrated window multi-anode MCP-PMT with better time performance and an integrated window structure will be more significant than that in the above study. In fact, we have already studied the processing technology and techniques for using $PbF_2$ instead of lead glass as the window material for our self-developed multi-anode MCP-PMT. However, due to the limitations of material properties such as the brittleness and deliquescence of $PbF_2$, there are still some challenges, but there is no theoretical barrier. Therefore, it is expected that in the near future, multi-anode MCP-PMTs using $PbF_2$ as the Cherenkov radiator window material will be successfully developed. At that time, based on the foundation laid by this research, experimental verification of the improvement in PET system performance and image quality brought about by PET imaging based on Cherenkov light detection will be conducted.

## 6. Conclusion

In this study, based on the detection of Cherenkov light, we used self-developed lead glass Cherenkov radiation integrator window multi-anode MCP-PMTs, combined with self-developed multi-channel electronics, to complete the development of PET detector modules, and built a dual-module PET imaging platform. Module-level calibration and correction were performed based on the characteristics of the platform. Based on this platform, a time resolution of 203.6 ps FWHM was obtained. Imaging experiments were completed for various radioactive sources and phantoms, and for the first time, experiments were performed to achieve module-level PET imaging based on the detection of Cherenkov light. In addition, we analyzed the improvement in imaging efficiency and system feasibility brought by multi-anode MCP-PMT, summarized the solutions to improve system uniformity and overcome the interaction between the ray and the MCP, and finally, based on this research, discussed the current problems with the use of lead glass as the Cherenkov radiator-integrated window material, revealing the significance of developing multi-anode MCP-PMT based on $PbF_2$-integrated window.

## Acknowledgements

This work was supported in part by the National Natural Science Foundation of China (Grant Nos. 12075267, 12275291, 12175265, 12205117) and Research Instrument and Equipment Development Project of Chinese Academy of Sciences (Grant No. YJKYYQ20200015)